\documentclass[11pt]{article}

\usepackage{harvard}
\usepackage{amsmath}
\usepackage{amssymb}
\usepackage{graphicx}
\usepackage{theorem}
\usepackage{times}

\begin{document}

%\begin{frontmatter}
\title{Two stock options at the races: Black-Scholes forecasts}

\author{G. Oshanin$^{1,2}$, G. Schehr$^3$\\
\\
$^1$ Physique Th\'eorique  de la Mati\`ere Condens\'ee, Universit\'e Pierre et Marie Curie, Paris, France\\
$^2$ Laboratory J.-V. Poncelet (UMI  CNRS 2615), Independent University of Moscow, Moscow, Russia\\
$^3$ Physique Th\'eorique, Universit\'e de Paris-Sud, France\\
}
%\thankstext{T1}{Footnote to the title with the `thankstext' command.}

\date{}
\maketitle

\begin{abstract}
Suppose one buys two very similar stocks and is curious about how much, after some time $T$,
one of them will contribute to the
overall asset, expecting, of course, that it should be around $1/2$ of the sum.
Here we examine this question within the
classical
Black and Scholes (BS) model, focusing on
the evolution of the probability
density function $P(w)$ of a random variable
 $w = a_T^{(1)}/(a_T^{(1)} + a_T^{(2)})$
where $a_T^{(1)}$ and $a_T^{(2)}$ are the values
of two (either European- or the Asian-style) options
produced by two absolutely identical BS stochastic equations.
We show that within the realm of the BS model
the behavior of $P(w)$ is surprisingly different from common-sense-based expectations.
For the European-style options
  $P(w)$ always undergoes a transition, (when $T$ approaches a certain threshold value), from a
unimodal to a bimodal form with the most probable values being
close to $0$ and $1$, and, strikingly, $w =1/2$
being the \textit{least} probable value.
This signifies that
the symmetry between two options spontaneously breaks
and just one of them
completely dominates the sum.
For path-dependent Asian-style options we observe
the same anomalous behavior, but only for a certain range of parameters.
Outside of this range, $P(w)$ is always a bell-shaped function with a maximum at $w = 1/2$.
\end{abstract}

%\begin{keyword}
%\kwd{Black-Scholes model}
%\kwd{European- and Asian-style options}
%\kwd{relative weight}
%\kwd{distribution function}
%\end{keyword}

%\end{frontmatter}

\section*{Introduction}

In finance, the style of an option is a general term denoting the
class into which the option falls, usually defined by the dates on which
the option may be exercised.
An Asian option is an option where the payoff is not determined
by the underlying price at maturity, contrary to European or American-style options, but
by the average underlying price over some pre-set period of time.
Such average-value
options are commonly traded on some Asian
(but also Western) markets being somehow
advantageous over European ones since the risk of asset
price manipulation near the maturity date is reduced due to their path dependence.

In classical
Black-Scholes settings \cite{r1,r2},
the underlying asset on which the Asian option is based is equal to  a stock
with price $S_t$ which follows the so-called geometric Brownian motion. In other words,
it is defined as  the
strong solution of the following linear stochastic differential equation:
\begin{equation}
d S_t = \omega \, S_t \, dt + \sigma \, S_t \, d B_t, \label{bs}
\end{equation}
where $\sigma$ is the volatility of $S_t$,
$B_t$ is a standard one-dimensional Brownian motion
and $\omega$ is a constant dependent on the nature of the asset:
it could be a stock, a currency, a commodity and etc.
For example, if $S_t$ is a stock paying a dividend at the continuous rate $\delta$,
one has $\omega = r - \delta$, where
$r$ is the risk-free rate of interest.

Resorting to It{\^o} calculus, one solves (\ref{bs}) to find
\begin{equation}
 S_t = S_0 \exp\left(- \frac{\sigma^2}{2} \mu t + \sigma B_t\right),
\end{equation}
where $S_0 = S_{t = 0}$ is the initial price and $\mu = 1 - 2 \omega/\sigma^2$ is a constant, which may
be positive, equal to zero or negative.
Further on, for continuous averaging with equal time-independent weights,
the random variable of interest - the value of an Asian option - is defined as the following functional of a Brownian
trajectory~\cite{r3,matsumoto_yor}:
\begin{equation}
\label{tau}
A_T = \frac{S_0}{T - T_0} \, \tau, \;\;\; \tau =  \int^T_{T_0} dt  \, \exp\left(- \frac{\sigma^2}{2} \mu t + \sigma B_t\right),
\end{equation}
where $T$ is the maturity date and $T_0$ is the time moment
when one starts to monitor the evolution of $S_t$.
Without lack of generality, we set $T_0 = 0$ in what follows.

In this paper we pose the following question which is, we believe, of a considerable conceptual interest:
Suppose that
one has not a single equation (\ref{bs}) but, say, two such
stochastic equations,
having the same volatility, the same $\omega$, the same
initial price $S_0$ and random noise terms, $d B_t^{(1)}$ and $d B_t^{(2)}$, which
have identic distributions.
These two equations produce
 two
identical European-style assets, $S_t^{(1)}$ and $S_t^{(2)}$,
which, in turn, generate two identical Asian-style random variables, $A_T^{(1)}$ and $A_T^{(2)}$.
What can be said about the distribution functions $P(W)$ and
$P(\mathcal{W})$ of random variables:
\begin{equation}
W = \frac{S_T^{(1)}}{S_T^{(1)} + S_T^{(2)}}, \label{W}
\end{equation}
and
\begin{equation}
\mathcal{W} = \frac{A_T^{(1)}}{A_T^{(1)} + A_T^{(2)}} = \frac{\tau_1}{\tau_1 + \tau_2}? \label{tauvar}
\end{equation}
These random variables describe a realization-dependent
relative weight of one asset in the sum of two assets.
Noticing that $S_t^{(1)}$ ($A_T^{(1)} $) and $S_t^{(2)}$ ($A_T^{(2)}$) are obviously
equal to each other on average, and moreover,
that all their higher moments are equal,
one might be tempted to
say that the distribution functions $P(W)$ and
$P(\mathcal{W})$ should  be bell-shaped functions with a maximum at $1/2$. They may broaden (initially both are delta-functions)
with growth of the maturity $T$, but still $W = 1/2$ and  $\mathcal{W} = 1/2$ should remain the most probable values.

Curiously, within the realm of the Black-Scholes model most often
this is not the case. We set out to show here
that the behavior of $P(W)$ and $P(\mathcal{W})$ is, in general,
surprisingly different from these common-sense-based expectations.

\section{Competition of two uncorrelated European-style variables.}

Note first that $W$ (and hence,
$P(W)$) is independent of $\mu$, since the factors
$\exp( - \sigma^2 \mu T/2)$ in the nominator and the denominator cancel each other.

%%%%%%%%%%%%%%%%%%%
\begin{figure} % figuur 1
%\fbox{\vbox to6pc{\hsize4cm\hfill\vfill}}
\centering
\includegraphics*{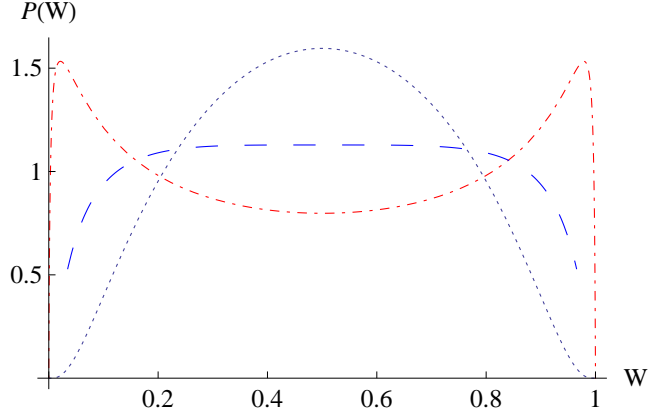}
\caption{Probability density $P(W)$ in (\ref{assets}) for different values of the effective maturity $\alpha_T$: Dotted curve corresponds to $\alpha_T = 0.25$, dashed - to $\alpha_T = 0.5$, while dot-dashed one - to $\alpha_T = 0.7$.}
\label{FIG1}
\end{figure}
%%%%%%%%%%%%%%%%%%%
For uncorrelated increments $d B_t^{(1)}$ and $d B_t^{(2)}$, the distribution function $P(W)$ can be
calculated exactly (see  Appendix \ref{euro}):
\begin{equation}
P(W) =  \frac{1}{\sqrt{8 \pi \alpha_T}} \frac{1}{W (1 - W)} \exp\left( - \frac{1}{8 \alpha_T} \ln^2\left(\frac{W}{1 - W}\right)\right), \label{assets}
\end{equation}
where $\alpha_T = \sigma^2 T/2$ is an effective maturity. This is the so-called logit-normal distribution, i.e., a probability distribution of a random variable whose logit has a normal distribution \cite{logit}.

Despite a relatively simple form, $P(W)$ in (\ref{assets}) contains a surprise:
 it shows a markedly different behavior
for $\alpha_T$ less or greater than $\alpha_T^c = 1/2$ (see Fig.~\ref{FIG1}).
For $\alpha_T < 1/2$ the distribution has a maximum at $W = 1/2$, which means that at early stages
both $S_T^{(1)}$ and  $S_T^{(2)}$ contribute proportionally;
it is thus most likely that each
variable defines just one half of the sum. However, when
$\alpha_T$ exceeds $1/2$,
$P(W)$ changes its shape from a unimodal
to a bimodal, $M$-shaped form with $P(W = 0) = P(W = 1) = 0$ and maximal values progressively closer to $0$ and $1$ as $\alpha_T \to \infty$.
Strikingly, $W = 1/2$ is now the \textit{least} probable value. It means that at sufficiently large
maturities the symmetry breaks  and one of the
 variables completely dominates the sum,
while the second one becomes a complete loser. Certainly, this is not the behavior one may expect on intuitive grounds.

\section{Competition of two uncorrelated Asian-style variables.}

Let $\Psi(\tau)$ denote the distribution function
of $\tau$ variables in (\ref{tauvar}). Then, for two uncorrelated Asian-style variables
 the probability
density $P(\mathcal{W})$ can be written as (see Appendix \ref{asian}):
\begin{equation}
P(\mathcal{W})
= \int^{\infty}_0 u \; du \; \Psi\left(\mathcal{W} u\right) \; \Psi\left((1 - \mathcal{W}) \; u\right). \label{dist}
\end{equation}
Clearly, $P(\mathcal{W})$ is symmetric around $\mathcal{W} = 1/2$. The question is
whether $\mathcal{W} = 1/2$ is always the maximum of the distribution?

The form of the distribution function $\Psi(\tau)$  was discussed
in the literature on mathematical finance
[see, e.g., \cite{r3,r4,r5}].
In addition, such $\tau$ variables appear in different domains of probability theory:
they are the continuous
counterparts of the so-called Kesten variables \cite{r6} that
play an important role in multiplicative stochastic processes and in the
renewal theory for products of random matrices.
In the physical literature, variables $\tau$ emerge in different contexts related to transport in disordered media.
In particular,
$\tau$ defines a resistance of a finite interval of
length $T$ offered to a passage \cite{r7} of particles
diffusing in
presence of a random, time-independent
 Gaussian
force with average value $\sigma^2 \mu/2$ and
variance $\sigma^2$ \cite{r8,r9,r10}.
Consequently,
inverse moments of $\tau$ define the moments of stationary currents
through the interval boundaries.
Within this context,  moments of $\tau$ \cite{r11,r12,r13,r14,r15} and the distribution function $\Psi(\tau)$ \cite{r14,r15} were also
calculated exactly.

For arbitrary $\mu$, $\Psi(\tau)$ is determined by \cite{r14,r15}:
\begin{equation}
\Psi(\tau) = \Psi_{con}(\tau) + \Psi_{dis}(\tau), \label{general}
\end{equation}
with
\[\Psi_{dis}(\tau) = \frac{\sigma^2}{2}  \exp\left(- \frac{1}{\tau'} \right)  \sum_{0 \leq n < \mu/2}  e^{- \alpha_T n (\mu-n)}
\frac{(-1)^n (\mu - 2 n)}{\Gamma\left(1 + \mu - n\right)} \left(\frac{1}{\tau'} \right)^{1+\mu-n} L_n^{\mu - 2n}\left(\frac{1}{\tau'}\right), \]
\label{discrete}
and
\[\Psi_{con}(\tau) = \frac{\sigma^2}{8 \pi^2 } \, \left(\frac{1}{\tau'}\right)^{(1 + \mu)/2} \, \int_0^{\infty} u \, du \, \left|\Gamma\left(- \frac{\mu}{2} + \frac{i u}{2}\right)\right|^2 W_{\frac{1+\mu}{2}, \frac{i u}{2}}\left(\frac{1}{\tau'}\right) \times \]
\[\times \exp\left(- \frac{\alpha_T}{4}\left(\mu^2 + u^2\right) -\frac{1}{2 \tau'}\right) \, {\rm sinh}(\pi u),\]
where $\tau' = \sigma^2 \tau/2$,
 $L^{\gamma}_n(x)$ are generalized Laguerre polynomials, $\Gamma(x)$ is the Gamma function, and $W_{\rho,\nu}(x)$ are Whittaker functions \cite{r16}. For $\mu > 0$
the distribution function in (\ref{general}) consists of discrete [$\Psi_{dis}(\tau)$]
and continuous branches [$\Psi_{con}(\tau)$],
while for $\mu \leq 0$ the distribution is determined by the continuous branch only.
Hence, unlike the distribution of $W$,  $P(\mathcal{W})$
will depend on the sign (and value) of $\mu$.  Consequently, we
will consider the cases of positive and negative $\mu$ separately.

\subsection*{$\mu = 0$.}

In this marginal case
$\tau$
defines an inverse
probability current in a finite Sinai chain \cite{r11,r12,r13}.
Occasionally, for this
case the distribution
$P(\mathcal{W})$
has been already calculated
in \cite{r17}, which studied the probability that a partially
melted heteropolymer at the melting temperature will either denaturate completely or return back
to a native helix state. Here
$\mathcal{W}$ defines the so-called splitting probability \cite{r7} -
the probability that the boundary between the helix and coil phases (which performs Sinai-type diffusion \cite{r9})
will first hit one of the extremities of the chain without
having ever reached the second extremity.

Adapting to our notations the result of \cite{r17}, we have
\begin{equation}
P(\mathcal{W}) =  \frac{1}{\pi \alpha_T \sqrt{\mathcal{W} (1 - \mathcal{W})^3}}
 \int_{-\infty}^{\infty} du\, \frac{{\rm cosh}(u)}{{\rm cosh}(\eta)} \cos\left(\frac{\pi u}{\alpha_T}\right) \times \label{dist3}
\end{equation}
\[ \times
\exp\left(\frac{\pi^2  - 4 u^2 -
4 \eta^2}{4 \alpha_T} \right), \, \eta = {\rm arcsinh}\left(\sqrt{\frac{\mathcal{W}}{1 - \mathcal{W}}} \, {\rm cosh}(u)\right).\]

%%%%%%%%%%%%%%%%%%%%%%
\begin{figure}
%\fbox{\vbox to6pc{\hsize4cm\hfill\vfill}}
\centering
\includegraphics*{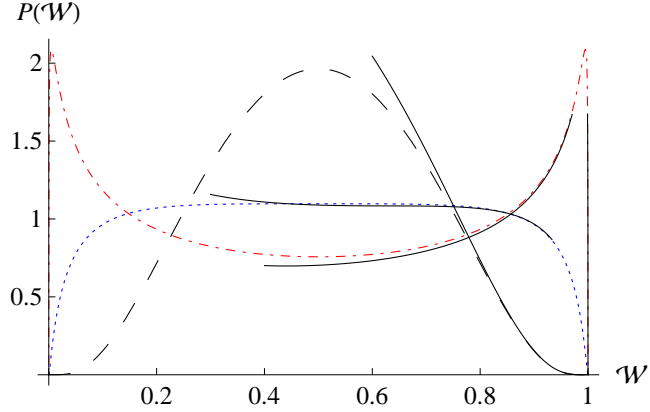}
\caption{Exact distribution $P(\mathcal{W})$, (\ref{dist3}),
for $\alpha_T=0.5$ (dashed curve), $\alpha_T=1.63$ (dotted), and
$\alpha_T=3.5$ (dash-dotted).  Thin solid lines are the corresponding asymptotic
results in (\ref{as1}) (see Appendix \ref{asian=0}).}
\label{PE-exact}
\end{figure}
%%%%%%%%%%%%%%%%%%%%%%

In Fig.~\ref{PE-exact} we plot $P(\mathcal{W})$  for several values
of an effective maturity $\alpha_T$. For sufficiently
low values of $\alpha_T$ the distribution is unimodal and centered
around $\mathcal{W} = 1/2$. Hence, in this early-time regime both
options  equally contribute to the total
asset. However, if we allow the options to ``mature``
longer,
we observe the same
surprising \textit{anomaly},
which we have already encountered in the previous section:
when $\alpha_T$ reaches a critical value $\alpha_T^c
\approx 1.63$, the maximum at $\mathcal{W} = 1/2$ ceases to exist
and the distribution becomes close to uniform for $0.2 \leq
\mathcal{W}\leq 0.8$.  Thus any value of $\mathcal{W}$ in this range
is nearly equally probable.
 For $\alpha_T$ exceeding $\alpha_T^c$, two maxima
emerge continuously and the distribution changes its shape becoming $M$-shaped bimodal,
with most probable values close to $0$ and $1$.
Although the \textit{average} value $\mathcal{W}$ is still  equal to $1/2$ in this regime,
the probability density now has a \textit{minimum} at  $\mathcal{W} = 1/2$.
Therefore, for $\mu = 0$ the path-dependence of the Asian-style variables
does not suppress the transition to the disproportionate behavior but only shifts it to later times;
$\alpha_T^c$ is
more than three times larger than the
corresponding value for the European-style variables.

\subsection*{$\mu > 0$.}

In this case
the distribution function in (\ref{general}) converges \cite{r14,r15}, as $\alpha_T \to \infty$, to a limiting form defined by the first term of a
discrete branch,
\begin{equation}
\label{discrete_first}
\Psi(\tau) \to \Psi_{\infty}(\tau) = \frac{2^{\mu} \sigma^2}{\Gamma(\mu)} \frac{1}{\left(\sigma^2 \tau\right)^{1 + \mu}} \, \exp\left(- \frac{2}{\sigma^2 \tau}\right).
\end{equation}
Hence, for $\alpha_T = \infty$ the variable $\tau$ has a very broad distribution characterized by a ''fat`` algebraic tail.
For $0 < \mu \leq 1$ this distribution is normalized but does not have moments.

We do not attempt
here
to present
an exact solution for $P(\mathcal{W})$ for arbitrary $\mu > 0$
and arbitrary
$\alpha_T$.
Instead, our aim will be to
get a conceptual understanding whether
$P(\mathcal{W})$ undergoes,  at a certain unknown
value of $\alpha_T$,
a transition from a unimodal
to a bimodal form.
This question
can be  immediately
answered if we find that
the corresponding limiting form of $P(\mathcal{W})$ exhibits such
a transition.

Plugging the limiting form in (\ref{discrete_first}) into (\ref{dist}) and performing integrations, we find that
in the limit $\alpha_T \to \infty$, the probability density $P(\mathcal{W})$ converges to a limiting form
\begin{equation}
P(\mathcal{W}) \to P_{\infty}(\mathcal{W}) = \frac{\Gamma(2 \mu)}{\Gamma^2(\mu)} \, \mathcal{W}^{\mu - 1} \left(1 - \mathcal{W}\right)^{\mu - 1}, \label{lim}
\end{equation}
i.e., it tends to a beta-distribution.

The distribution on the right-hand-side of (\ref{lim}) has a different shape (modality)
for $0 < \mu < 1$, $\mu = 1$ or $\mu > 1$. For $0 < \mu < 1$ the distribution
$P_{\infty}(\mathcal{W})$
is a $U$-shaped bimodal.
For $\mu = 1$ it is uniform.
Finally, $P_{\infty}(\mathcal{W})$ is  unimodal,  centered at $ \mathcal{W} = 1/2$ for $\mu > 1$.

Therefore, for $0 < \mu < 1$ the distribution $P(\mathcal{W})$ will
change its shape from a unimodal to a bimodal at a certain value $\alpha_T^c$.
For the marginal $\mu = 1$ case, the distribution $P(\mathcal{W})$ will tend
to a uniform distribution as $\alpha_T \to \infty$. Finally, for $\mu > 1$,
the distribution will always remain unimodal and centered around the most
probable value $\mathcal{W} = 1/2$. Note that a similar transition was observed  for the distribution
of the occupation time on the positive axis for Sinai model with a drift $\mu$ ~\cite{satya_prl,sanjib_pre}. Interestingly, this transition also occurs for $\mu = 1$.

\subsection*{$\mu < 0$.}

In this case the distribution $\Psi(\tau)$ is given by the continuous branch in (\ref{general}).
Note that the expression in (\ref{general}) can not be used
directly, since
it does not allow one to perform an integration over $\tau$ in (\ref{dist}). Thus,
we will first  try
to obtain a plausible approximation for $\Psi(\tau)$ valid for large $\alpha_T$, which will allow
us to perform the integration over $\tau$.
Then,
on base of this result we
will see whether the transition to the disproportionate behavior indeed takes place or not.
After some straightforward
manipulations (see Appendix \ref{asian<0}), we find
\begin{equation}
\Psi(\tau) \sim C \, \frac{\exp(-1/\tau')}{\tau'^{1 + \mu/2}} \, U\left(-\frac{\mu}{2},1,\frac{1}{\tau'}\right) \, \exp\left( - \frac{{\rm arcsinh}^2(\sqrt{\tau'})}{\alpha_T}\right), \label{app}
\end{equation}
where $C$ is a constant (see Appendix \ref{asian<0}) and
$U(a,b,z)$ is the confluent hypergeometric function \cite{r16}.
Note that for sufficiently large $\alpha_T$ the approximation in (\ref{app})
works fairly well for any value of $\tau$
 (see Fig. ~\ref{fig_comparison_psi} in Appendix \ref{asian<0}). Moreover, it exhibits exact asymptotic behaviors
 in the limits $\tau \to 0$ and $\tau \to \infty$.

Equation (\ref{app})
predicts that $\Psi(\tau)$ has a log-normal tail
as $\tau \to \infty$.  Given such a slow decay, one expects that the integral in (\ref{dist})
is dominated by
large values of $u$. One finds then that, for sufficiently
large $\alpha_T$, and $\mathcal{W}$ not too close to $0$ or $1$,
\begin{equation}
 P({\mathcal{ W}}) \sim \frac{1}{\sqrt{\alpha_T}} \frac{1}{W(1-W)}, \label{approx_m}
\end{equation}
while for $W \to 0$ or $1$ one finds that $P({\cal W})$ obeys (\ref{assets}) (with possible logarithmic corrections, see Appendix \ref{asian=0}, (\ref{as1})).
Therefore,  for $\mu < 0$ the
distribution $P({\cal W})$ does not reach a limiting form when $\alpha_T \to \infty$.
This means, in turn, that \textit{for all} $\mu < 0$, the distribution $P({\mathcal W})$
exhibits a generic transition from a bell-shaped to an $M$-shaped form, as $\alpha_T$ passes through some
critical value $\alpha_T^c$.

In fact, this is a rather counter-intuitive result. The random variables $A_T$ are integrals of a geometric Brownian motion - an exponential of a symmetric Brownian motion plus a constant drift term $- \sigma^2 \mu/2$. One may naturally expect that for sufficiently large $\mu$ a contribution due to symmetric Brownian motion will be insignificant. This is precisely the behavior we observed in the $\mu > 0$ case, for which the transition to the disproportionate behavior takes place only for $\mu \in [0,1[$ and is absent for $\mu > 1$. Surprisingly, this is not the case for $\mu < 0$ and "disorder" embodied in Brownian terms $B_t$ turns out to be always relevant, despite the fact that the constant drift term clearly provides a dominant contribution (for sufficiently large $\mu$) in the exponential.

\begin{figure}
\centering
\includegraphics[width=0.6\linewidth]{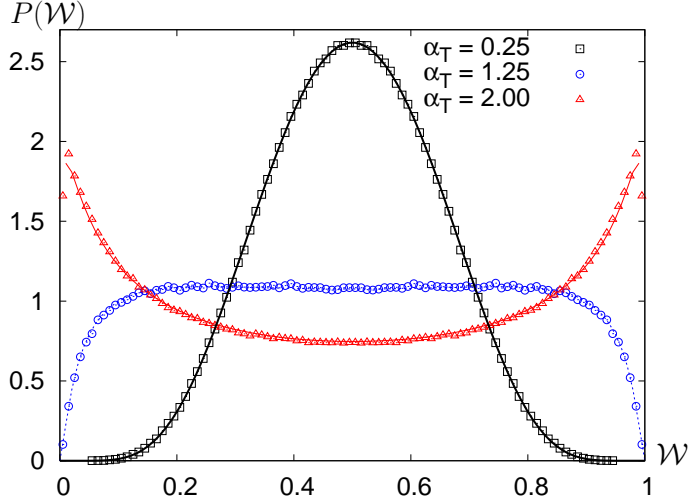}
\caption{Probability density ${P}({\cal W})$ obtained from numerical simulations (symbols)
of the race between two uncorrelated Asian-style variables for
for $\mu = -1.7$ and
for two different
maturities $\alpha_T = 0.25$ (a bell-shaped curve) and $\alpha_T = 1.5$ (an $M$-shaped curve). The critical value $\alpha_T^c \simeq 1.12$.
The dashed lines correspond to a fit according to (\ref{assets}) with $\alpha_T$ replaced with a fitting parameter $\tilde \alpha_T$.
\label{fig_muneg_uncorrel}}
\end{figure}

We have performed numerical simulations of the race between two uncorrelated Asian-style random variables [see (\ref{rw}, \ref{tau_num}) below] which confirm our conclusions on the transition to a disproportionate behavior for any $\mu < 0$. In Fig. \ref{fig_muneg_uncorrel} we plot $P({\cal W})$ for $\mu = -1.7$ for three different values of the maturity $\alpha_T$, which clearly
shows a transition from a unimodal to an $M$-shaped form, as the maturity exceeds
a critical value $\alpha_T^c \simeq 1.12$.
Guided by the above analytic argument (\ref{approx_m}), we have fitted these distributions using the expression
for the European-style variable, $P(W)$ in (\ref{assets}), in which we replaced $\alpha_T$ by
  some effective maturity $\tilde \alpha_T$ used as a fitting parameter.
  As one may notice,
the quality of the fit  in Fig. \ref{fig_muneg_uncorrel} is very good.
We also found numerically that $\alpha_T^c \equiv \alpha_T^c(\mu)$ is a
slowly decreasing function of $\mu$ for $\mu < 0$, which implies that the larger, (by absolute value) $\mu$ is, the earlier the transition takes place.
 %%%%%%%%%%%%%%%%%%%
%%%%%%%%%%%%%%%%%%%

\section{Correlated increments.}

So far we have concentrated on the case of independent variables, to which one may
object claiming that such a transition is spurious and
correlations \cite{r18} between the increments $dB_t^{(1)}$ and $dB_t^{(2)}$ will "stabilize" the behavior of the European- and/or the Asian-style variables. In the remainder, we proceed to show that the transition to the bimodal shape is robust and the only effect of correlations is to shift the transition to later times.

We focus on the case of very strongly correlated increments,
when the penalty for having different $dB_t^{(1)}$ and  $dB_t^{(2)}$
grows with $ u_t = dB_t^{(1)} - dB_t^{(2)}$ as $u_t^2$. In this case
the distribution function of the increments reads:
\begin{eqnarray}
F\left(dB_t^{(1)},dB_t^{(2)}\right) &=& \frac{\sqrt{2 + \chi^2}}{2 \pi \chi dt} \exp\left( - \frac{\left(dB_t^{(1)}\right)^2}{2 dt} - \frac{\left(dB_t^{(2)}\right)^2}{2 dt}\right) \nonumber \\
&\times& \exp\left(- \frac{\left(dB_t^{(1)} - dB_t^{(2)}\right)^2}{2 \chi^2 dt}\right)
\label{gauss}
\end{eqnarray}
Physically, it corresponds to the
situation of two Brownian particles coupled by a Hookean spring with rigidity $1/\chi$.
The parameter $\chi$ sets the scale of correlations.
When $\chi \to 0$ (an infinite rigidity of the spring), the function $F$ in (\ref{gauss}) converges to a delta-function which
signifies that the increments are forced to be equal to each other, $dB_t^{(1)} = dB_t^{(2)}$.
Conversely, when $\chi$ is large, the function $F$ in (\ref{gauss}) will tolerate
large deviations of $dB_t^{(2)}$ from  $dB_t^{(1)}$. For $\chi = \infty$ we recover the limit of independent
variables.

For $F$ in (\ref{gauss}),
the distribution function $P(W)$ of the random variable $W$ can be calculated exactly (see Appendix \ref{corr}):
\begin{equation}
\label{Gauss}
P(W) =  \left(\frac{2 + \chi^2}{8 \pi \chi^2 \alpha_T}\right)^{1/2} \frac{1}{W (1 - W)} \exp\left( - \frac{2 + \chi^2}{8 \chi^2 \alpha_T} \ln^2\left(\frac{W}{1 - W}\right)\right),
\end{equation}
and appears to have essentially the same form as the result for
two \textit{uncorrelated} European-style variables, (\ref{assets}).
The only difference is that now $\alpha_T$ is renormalized by $\chi$, so that the
transition to the bimodal form occurs at
$\alpha_T^c = (2 + \chi^2)/2 \chi^2 \geq 1/2$.
Hence, in presence of correlations the transition to the bimodal form is postponed for later times.

\begin{figure}[ht]
\begin{center}
\includegraphics[width=0.7 \linewidth]{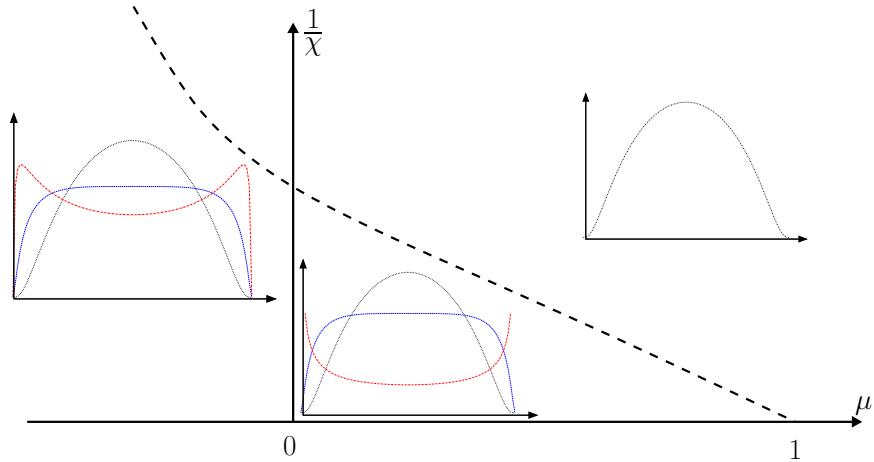}
\caption{A sketch of the phase diagram in the $(\mu,1/\chi)$ plane. The dotted line starting at $\mu = 1$ for $1/\chi = 0$ is the critical line $\mu_c(\chi)$. For $\mu > \mu_c$, the distribution $P({\cal W})$ remains unimodal for all maturities. For $\mu < \mu_c$ we depict
 the shape of $P({\cal W})$ for $\alpha_T < \alpha_T^c$, $\alpha_T \simeq \alpha_T^c$ and $\alpha_T > \alpha_T^c$. For $0 < \mu < \mu_c$, the distribution
 $P({\cal W})$ evolves from a bell-shaped curve to a $U$-shaped one, while for $\mu < 0$, $P({\cal W})$ evolves from a bell-shaped curve to an $M$-shaped one as $\alpha_T$ crosses $\alpha^c_T$.}\label{phase_diag}
\end{center}
\end{figure}

For correlated Asian options,
analytic calculations are much more involved and here we again resort to numerical analysis. To this purpose, we first generate two random walks $x_n^{(1)}$ and $x_n^{(2)}$ which evolve according to the following rules (in discrete time):
\begin{eqnarray}\label{rw}
&&x_0^{(1)} = x_0^{(2)} = 0 \;, \nonumber \\
&&x_n^{(1)} = x_{n-1}^{(1)} - \frac{\mu}{2} \sigma^2 + \eta_n^{1} \;, \; x_n^{(2)} = x_{n-1}^{(2)} - \frac{\mu}{2} \sigma^2 + \eta_n^{(2)} \;,\;n \geq 1 \;,
\end{eqnarray}
where $\eta_n^{(i)} \equiv dB_t^{(i)}$, $i=1,2$ and $\sigma^2 = E\left([dB_t^{(i)}]^2\right) = {(1+\chi^2)}/{(2+\chi^2)}$. One has, of course, $E\left(\eta^{(i)}_n \eta_{n'}^{(j)}\right) = 0$ if $n \neq n'$.  On the other hand, it is easy to see from (\ref{gauss}) that the random variables $u_n = \eta^{(2)}_n - \eta^{(1)}_n$ and $v_n=\eta^{(2)}_n + \eta^{(1)}_n$ are independent Gaussian random variables with correlations
\begin{eqnarray}\label{un_vn}
E(u_n^2) = \frac{2 \chi^2}{2 + \chi^2} \;, \; E(v_n^2) = 2 \;.
\end{eqnarray}
Therefore, in order to generate the random variables $\eta_n^{(1)}$ and $\eta_n^{(2)}$ -- which in the limit of the large number of steps converge to $dB_t^{(1)}$ and $dB_t^{(2)}$ distributed according to (\ref{gauss}), -- we generate two independent Gaussian random variables $u_n$ and $v_n$ satisfying (\ref{un_vn}), from which we obtain the desired $\eta_n^{(1)} = (v_n-u_n)/2$  and $\eta_n^{(2)} = (v_n+u_n)/2$.  Finally, from these two random walks it is straightforward to obtain two correlated Asian style options $\tau_i$'s from (\ref{tau}):
\begin{eqnarray}\label{tau_num}
\tau_i = \exp{\left[\sum_{n=0}^N x_n^{(i)} \right]} \;,
\end{eqnarray}
where $N \propto T$.  One can then compute the distribution $P({\cal W})$. By varying the parameters $\chi$ and $\mu$ we obtain the phase diagram which is depicted in Fig. \ref{phase_diag}. This is achieved as follows:
\begin{itemize}
\item[$\bullet$] {For $\mu > 1$, we have shown previously that even in the absence of correlations, (which corresponds to
$\chi \to \infty$), the distribution $P({\cal W})$ is always unimodal so that it will remain unimodal for any finite value of $\chi$. In this case we have checked numerically that the distribution $P({\cal W})$ converges to a $N$-independent, unimodal curve for $N$ in (\ref{tau_num}) sufficiently large.}
\item[$\bullet$]{For $0 < \mu < 1$ and $\chi \to \infty$, we have shown  that the shape of $P({\cal W})$ changes from a bell-shaped to a $U$-shaped form as $N \propto \alpha_T$  passes through a critical value $N_c(\mu, \chi \to \infty)$, which we have numerically verified. For finite  but sufficiently large $\chi$, $\chi > \chi_c(\mu)$ we continue to observe such a transition as $N$ crosses a critical value $N_c(\mu,\chi)$, where $N_c(\mu, \chi)$ increases as $\chi$ is decreased, and eventually $N_c(\mu, \chi)$ diverges when $\chi \to \chi_c(\mu)$. For $\chi < \chi_c(\mu)$ one observes instead that  $P({\cal W})$ converges to a $N$-independent unimodal distribution. In our simulation, we thus identify $\chi_c(\mu)$ as the largest value of $\chi$ for which, for a given $0 < \mu < 1$ fixed, such a convergence is observed. We have checked numerically that the "critical region" is very narrow, in the sense that $N_c(\mu, \chi)$ grows very rapidly when $\chi \to \chi_c$, which thus allows for a reliable estimate of the critical line depicted as a dotted line in Fig. \ref{phase_diag} in the $(\mu,1/\chi)$ plane. Note, however, that a precise estimate of this critical line would certainly require a more careful analysis.
}
\item[$\bullet$]{For $\mu < 0$ we observe instead a transition from an $M$-shaped to a bell-shaped form of the distribution but the effect of correlations is qualitatively similar to the case $0 < \mu < 1$. The critical line is identified as described above for $0 < \mu < 1$.}
\end{itemize}

%%%%%%%%%%%%%%%%%%%%%%
%%%%%%%%%%%%%%%%%%%%%%%%%%%%%%%%%%%%%
This phase diagram shown in Fig. \ref{phase_diag} can be summarized as follows. For finite $\chi$, we find that $P({\cal W})$ always undergoes a transition from a unimodal to a bimodal form as the maturity passes through a critical value $\alpha_T^c$ for all $\mu < \mu_c\equiv \mu_c(\chi)$. As we expected,
$\mu_c$ is a decreasing function of the strength of correlations: this is shown
by the dotted line in Fig. \ref{phase_diag}. Similarly to the uncorrelated case,
one also finds a different behavior for $0 < \mu < \mu_c $ and $\mu < 0$: in the first
case, $P({\cal W})$ changes from a bell-shaped  to a $U$-shaped form, while for $\mu < 0$ it changes from a bell-shaped  to an $M$-shaped form. This is depicted in Fig.
\ref{phase_diag}.
Finally, for $\mu < \mu_c$, we observe that $\alpha_T^c$ is an increasing function of the strength of correlations, similarly to the case of the European-style variables.

We conclude with several words
concerning a common feature of the transition to the disproportionate behavior observed in our paper.
As one may notice, such a transition always takes place in situations when the first moment of the distribution function of the underlying $S$ or $A$ diverges as $T \to \infty$, and does not take place if the first moment remains finite at $T = \infty$. It is a bit intriguing to see how the behavior at $T = \infty$ defines the transition which
 takes place at a finite $T$.

\appendix

\section{Independent European-style variables}\label{euro}

Let $E\left\{\exp\left(- \lambda W\right)\right\}$, $\lambda \geq 0$, denote the moment generating function of the random variable $W$. The curly brackets here and henceforth denote averaging with respect to the distributions of Brownian motions $B_t^{(1)}$ and $B_t^{(2)}$.
Explicitly, this function can be written down as
\[
E\left\{e^{- \lambda W}\right\} = \frac{1}{2 \pi T} \int_{- \infty}^{\infty} \int_{- \infty}^{\infty} dB_1 \, dB_2 \, \exp\left(- \frac{B_1^2 + B_2^2}{2 T} - \lambda \frac{e^{\sigma B_1}}{e^{\sigma B_1} + e^{\sigma B_2}}\right).
\]
Integrating over $d B_1$, we formally change the integration variable $B_1 \to W$, which yields
\[
E\left\{e^{- \lambda W}\right\} =  \frac{1}{\sqrt{8 \pi \alpha_T}} \int_{0}^{1} \frac{d W}{W (1 - W)} \exp\left( - \lambda W - \frac{1}{8 \alpha_T} \ln^2\left(\frac{W}{1 - W}\right)\right),
\]
from which one immediately deduces (\ref{assets}).

\section{Independent Asian-style variables}\label{asian}

The moment generating function of $\mathcal{W}$ can be written as
\[
E\left\{ e^{- \lambda \mathcal{W}}\right\} = \int^{\infty}_0 \,  d\tau_1 \, \int^{\infty}_0 \, d\tau_2 \, \Psi(\tau_1) \, \Psi(\tau_2) \, \exp\left(-\lambda \frac{\tau_1}{\tau_1 + \tau_2}  \right).
\]
Integrating over $d \tau_2$, we formally change the integration variable
from $\tau_2$ to $\mathcal{W}$ to give
\[
E\left\{ e^{- \lambda \mathcal{W}} \right\} = \int^1_0 \, d \mathcal{W} \,  e^{- \lambda \mathcal{W}} \int^{\infty}_0 \, \tau_1 \, d\tau_1 \, \Psi(\mathcal{W} \tau_1) \, \Psi\left((1 - \mathcal{W}) \tau_1\right),
\]
from which one may read off the result in (\ref{dist}).

\subsection*{$\mu>0$}\label{asian=0}

We note that in the limit
$\mathcal{W} \to 1$, $P(\mathcal{W})$ in (\ref{assets}) has the following asymptotic representation \cite{r17}:
\begin{equation}
\label{as1}
P(\mathcal{W})  \sim \frac{\left(1 + \sqrt{\mathcal{W}}
\ln z\right)^{-1/2}}{\sqrt{\pi \alpha_T \mathcal{W}}\,\, (1 - \mathcal{W})}
\exp\!\left(- \frac{\ln^2 z}{\alpha_T}
\! +\! \frac{\pi^2 \sqrt{\mathcal{W}} \ln z}
{4 \alpha_T \left(1 + \sqrt{\mathcal{W}} \ln z\right)}\right)\!,
\end{equation}
where $z = (1 + \sqrt{\mathcal{W}})/\sqrt{(1 -\mathcal{W} )}$.  This asymptotic form agrees quite
well with the exact result in (\ref{dist3}), not only when $\mathcal{W}\to 1$,
but also for moderate values of $\mathcal{W}$ (Fig.~\ref{PE-exact}).
The asymptotics of $\mathcal{P}(\mathcal{W})$ for $\mathcal{W} \to 0$ can be obtained by merely
changing $\mathcal{W}$ to $1 - \mathcal{W}$.

It may be also worthy to
 remark that in the limit $\mathcal{W} \to 1$ the asymptotic form in (\ref{as1})
follows, apart of a logarithmic factor $\ln^{1/2}\left(1/(1 - \mathcal{W})\right)$,
the asymptotic form of the parental distribution $\Psi(\tau)$ in
(\ref{general}) \cite{r13,r14,r15}:
\[
 \Psi(\tau) \sim  \frac{1}{2 \sqrt{\pi \alpha_T}} \frac{1}{\tau} \exp\left( - \frac{1}{4 \alpha_T} \ln^2\left(\tau\right)\right), \;\;\; \tau \to \infty
\]
For fixed $\mathcal{W}$, the probability density
$P(\mathcal{W}) \sim 1/\sqrt{\alpha_T}$ when $\alpha_T \to \infty$, which signifies
that the large-$T$ behavior of $P(\mathcal{W})$ in (\ref{as1}) is supported by \textit{negative} moments of $\tau$, which decay as $E(1/\tau^n) \sim 1/\sqrt{\alpha_T}$ regardless of the order $n$ \cite{r13,r14,r15}.

\subsection*{$\mu < 0$}\label{asian<0}

Whittaker function has the following integral representation:
\vspace{0.1in}
\[\left|\Gamma\left(- \frac{\mu}{2} + \frac{i u}{2}\right)\right|^2 W_{\frac{1+\mu}{2}, \frac{i u}{2}}\left(y\right) = \]
\[= 2^{\mu + 2} \, y^{(\mu + 1)/2} \exp{\left(- \frac{y}{2}\right)}
\int_0^\infty \frac{dx}{ x^{\mu +1}} \,  \exp{\left(-\frac{x^2}{4 y} \right)} {\mathrm K}_{iu}(x),\]
\vspace{0.1in}
where ${\mathrm K}_{iu}(x)$ is the modified Bessel function \cite{r16}. Using the latter equation and a standard integral representation of ${\mathrm K}_{iu}(x)$,
\vspace{0.1in}
\[{\mathrm K}_{iu}(x) = \int^{\infty}_0 dt \, \cos(u t) \exp\left(- x \, {\rm cosh}(t)\right),\]
\vspace{0.1in}
we find
\vspace{0.1in}
\begin{equation}\label{factor_psi}
 \Psi(\tau) = \frac{2^{\mu - 1} \sigma^2 \exp{(-\alpha_T \mu^2/4)}}{\pi^2} \, \frac{\exp(-1/\tau')}{\tau'^{\mu+1}} \, \psi(\alpha_T,\tau'),
\end{equation}
where
$\tau' = \sigma^2 \tau/2$ and $\psi(\alpha_T,\tau)$ is given by
\vspace{0.1in}
\begin{equation}\label{start_psi}
\psi(\alpha_T, \tau) = \int_0^\infty \frac{dx}{x^{1 + \mu}}
\exp\left(-\frac{\sigma^2 \tau x^2}{8} \right)
\theta\left(x,\frac{\alpha_T}{2} \right),
\end{equation}
with
\begin{equation}\label{def_theta}
\theta\left(x,\frac{\alpha_T}{2} \right) =
\frac{2e^{\frac{\pi^2}{\alpha_T}}}{(\pi \alpha_T)^{\frac{3}{2}}} \int_0^\infty
d\xi \exp{\left(- x \cosh{(\xi)} - \frac{\xi^2}{\alpha_T}\right)} \times
\end{equation}
\[
\times \left(\pi\cos{\left(\frac{2\pi \xi}{\alpha_T}\right)}- \xi \sin{\left(\frac{2\pi \xi}{\alpha_T}\right)}  \right).\]
We now focus on $\psi(\alpha_T,\tau)$ seeking for a plausible approximation in the limit $\alpha_T \gg 1$.
Note first that  $\psi(\alpha_T,\tau)$  is a monotonically decreasing function of $\tau$ and
\begin{equation}\label{small_tau}
\psi(\alpha_T, \tau =0) = \frac{\Gamma^2(-\mu/2)}{\sqrt{\pi} 2^{1+\mu}
  \alpha_T^{3/2}} + o\left(  \alpha_T^{-3/2}\right).
\end{equation}
In the large $\tau$ limit, the integral in
(\ref{start_psi}) is dominated by the small $x$ behavior of the
integrand. To obtain the small $x$
behavior of $\theta\left(x,\frac{\alpha_T}{2} \right)$ in the
large $\alpha_T$ limit, we take
$x = 1/(2 \sinh{(y \sqrt{\alpha_T})})$ and also
change the integration variable $\xi = u \sqrt{\alpha_T}$ in (\ref{def_theta}).
Setting then $\alpha_T \to \infty$ while keeping $y$ fixed, we have
\begin{equation}\label{theta_3}
\theta\left(x,\frac{\alpha_T}{2} \right) \simeq \frac{2}{\sqrt{\pi}
  \alpha_T} y e^{-y^2} \equiv
\end{equation}
\[  \equiv  \frac{2}{\sqrt{\pi}
  \alpha_T^{3/2}} \, \mathrm{arcsinh}{\left(\frac{1}{2 x}\right)} \, \exp{\left(-\frac{1}{\alpha_T} \mathrm{arcsinh}^2\left(\frac{1}{2 x}\right)\right)},
\]
which defines the small $x$ asymptotic behavior of $\theta(x,\alpha_T/2)$.

On the other hand, for large $x$
\[\theta\left(x,\frac{\alpha_T}{2} \right) \sim 2 K_0(x)/\sqrt{\pi}\alpha^{3/2}.\]
Hence, we may approximate $\theta\left(x,\frac{\alpha_T}{2} \right)$,  for  sufficiently large $\alpha_T$, as
\begin{equation}\label{approx_theta}
 \theta\left(x,\frac{\alpha_T}{2} \right) \simeq \frac{2}{\sqrt{\pi}\alpha_T^{3/2}} \, K_0(x) \, \exp{\left(-\frac{1}{\alpha_T} \mathrm{arcsinh}^2\left(\frac{1}{2 x}\right)  \right)}.
\end{equation}
Note that this approximate form
reproduces correctly
the exact behavior of $\theta\left(x,\frac{\alpha_T}{2} \right)$ both for $x \to 0$ (\ref{theta_3}) and $x \to \infty$.

 We use next the small $x$ asymptotic behavior of $\theta\left(x,\frac{\alpha_T}{2} \right)$, (\ref{theta_3}), to obtain from (\ref{start_psi})
the following large-$\tau$ asymptotic of $\psi(\alpha_T,\tau)$:
\begin{equation}\label{large_tau}
\psi(\alpha_T,\tau) \simeq \frac{\Gamma(-\mu/2)}{\sqrt{\pi}
  \alpha_T^{3/2} 2^{1+\mu}} \tau'^{\mu/2} \log{(\tau')} \exp{\left(-
  \frac{\log^2{\tau'}}{4 \alpha_T} \right)}.
\end{equation}

The small-$\tau$ asymptotic behavior of $\psi(\alpha_T,\tau)$
can be deduced from the large-$x$ asymptotic of
$\theta\left(x,\frac{\alpha_T}{2} \right)$, which yields
\begin{eqnarray}\label{approx_moderate}
 \psi(\alpha_T, \tau) \simeq \frac{\Gamma^2(-\mu/2)}{2^{\mu + 1} \sqrt{\pi} \alpha_T^{3/2}} \, \tau'^{\mu/2} \, U\left(-\frac{\mu}{2}, 1,\frac{2}{\sigma^2 \tau}\right),
\end{eqnarray}
where $U(a,b,z)$ is the confluent hypergeometric function \cite{r16}.

Combining the estimates in (\ref{large_tau}) and (\ref{approx_moderate}), we obtain
 the approximate form for $\Psi(\tau)$ in (\ref{app}) with
\[C = \frac{\sigma^2 \Gamma^2(-\mu/2)}{4 \pi^{5/2} \alpha_T^{3/2}} \exp\left(- \frac{\alpha_T \mu^2}{4}\right).
\]

\begin{figure}[h]
\begin{center}
\includegraphics{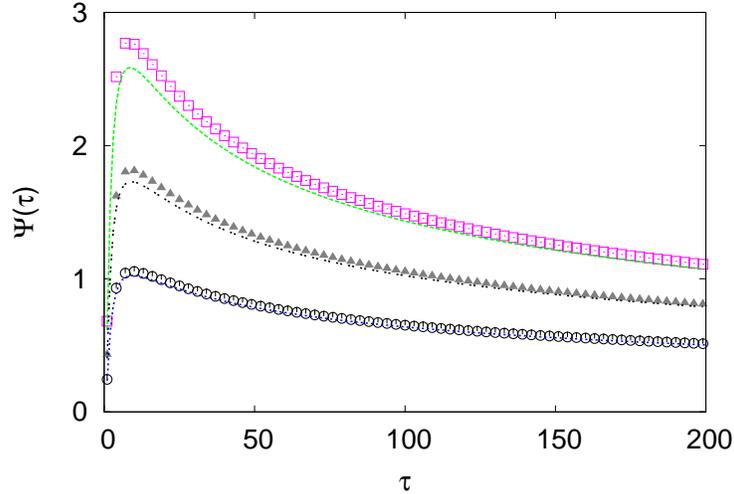}
\caption{Probability density $\Psi(\tau)$ for $\mu=-1$ and $\alpha_T = 20, 30, 50$ (from top to bottom). Each pair of curves have been rescaled by an appropriate factor
so that they can be shown on a same graph. The symbols correspond to the exact value of $\Psi(\tau)$, obtained by a numerical evaluation of the integrals in (\ref{factor_psi}) and (\ref{start_psi}), while the solid lines correspond to the approximation given in (\ref{app}).}
\label{fig_comparison_psi}
\end{center}
\end{figure}

In Fig. \ref{fig_comparison_psi}, we compare our approximate expression for $\Psi(\tau)$ in (\ref{app}) against the numerical evaluations of the integrals in (\ref{factor_psi}) and (\ref{start_psi}) for $\mu=-1$ and different values of $\alpha_T = 20, 30, 50$.
One notices that our approximation is quite accurate for all $\tau >0$.
As expected, the approximation works
better as $\alpha_T$ increases.

\section{Correlated increments}\label{corr}

 We first
divide the interval $[0,T]$ into $N$ subintervals $dt$
approximating the Brownian trajectories $B_T^{(1)}$ and
$B_T^{(2)}$ by the trajectories of
random walks in discrete time
(with time-step $dt$) $k$:
\[
B_T^{(n)} = \sum_{k = 1}^N  dB_k^{(n)}, \, n=1,2,
\]
where $dB_k^{(n)}$ are the values of the
increments at time moment $k$.
In what follows, we will be interested by the behavior in the
limit $N \to \infty, dt \to 0$ with
$N dt = T$ kept fixed.

Then, a random variable $W$, which defines a contribution of a given European-style
option into the sum of two such options, can be written formally as
\[
W = \left(1 + \prod_{k=1}^N \exp\left(\sigma (dB_k^{(1)} - dB_k^{(2)}) \right)\right)^{-1}.
\]
Then, for the generating function of this random variable we have
\[
E\left\{e^{- \lambda W}\right\} = \left(\frac{\sqrt{2 + \chi^2}}{2 \pi \chi dt}\right)^N \int^{\infty}_{-\infty} \ldots \int^{\infty}_{-\infty} \prod^N_{k=1} dB_k^{(1)} dB_k^{(2)} \times
\]
\[ \times \exp\left(- \frac{\left(dB_k^{(1)}\right)^2}{2 dt} - \frac{\left(dB_k^{(2)}\right)^2}{2 dt} - \frac{\left(dB_k^{(2)} - dB_k^{(2)}\right)^2}{2 \chi^2 dt}\right) \times \]
\[ \times \exp\left(-\frac{\lambda}{1 + \prod_{k=1}^N \exp\left(\sigma \left(dB_k^{(2)} - dB_k^{(2)}\right)\right)}\right) = \]
\[= \left(\frac{\sqrt{2 + \chi^2}}{2 \pi \chi dt}\right)^N \int^{\infty}_{-\infty} \ldots \int^{\infty}_{-\infty} \prod^N_{k=1} dB_k^{(1)} du_k \times \]
\[ \times \exp\left(- \frac{\left(dB_k^{(1)}\right)^2}{dt} - \frac{dB_k^{(1)}}{dt} u_k - \frac{(1 + \chi^2) u_k^2}{2 \chi^2 dt}\right) \times \]
\[\times \exp\left(-\frac{\lambda}{1 + \prod_{k=1}^N \exp\left(\sigma u_k\right)}\right),\]
which reduces, upon the integration over $dB_k^{(1)}$, to
\[
E\left\{e^{- \lambda W}\right\} = \frac{1}{\left(2 \pi g^2 dt\right)^{N/2}} \int^{\infty}_{-\infty} \ldots \int^{\infty}_{-\infty} \prod^N_{k=1} du_k \times \]
\[
\times \exp\left(- \frac{u_k^2}{2 g^2 dt}\right) \exp\left(-\frac{\lambda}{1 + \prod_{k=1}^N \exp\left(\sigma u_k\right)}\right),
\]
where $g^2 = 2 \chi^2/(2 + \chi^2)$.

Further on, performing the integration over $du_N$ we formally change the
integration variable $u_N \to W$, which yields
\[
E\left\{e^{- \lambda W}\right\} = \frac{1}{\sigma \left(2 \pi g^2 dt\right)^{N/2}} \int^1_0 \frac{dW}{W (1 - W)} e^{- \lambda W}  \int^{\infty}_{-\infty} \ldots \int^{\infty}_{-\infty} \prod^{N-1}_{k=1} du_k \times \]
\[ \times \exp\left(- \frac{u_k^2}{2 g^2 dt}\right) \exp\left(-\frac{1}{2 g^2 dt}\left(
\frac{1}{\sigma} \ln\left(\frac{1 - W}{W}\right) - \sum_{k=1}^{N - 1} u_k\right)^2
\right).
\]
The latter equation implies that the distribution function
$P(W)$ in case of two European-style variables with correlated increments
is given by
\[
P(W) =  \frac{1}{\sigma \left(2 \pi g^2 dt\right)^{N/2}} \frac{1}{W (1 - W)}\int^{\infty}_{-\infty} \ldots \int^{\infty}_{-\infty} \prod^{N-1}_{k=1} du_k \, \exp\left(- \frac{u_k^2}{2 g^2 dt}\right) \times \]
\[ \times \exp\left(-\frac{1}{2 g^2 dt}\left(
\frac{1}{\sigma} \ln\left(\frac{1 - W}{W}\right) - \sum_{k=1}^{N - 1} u_k\right)^2
\right)
\]
Finally, by taking the advantage of the integral identity
\[
\frac{\exp\left(- u^2/2 A\right)}{\sqrt{2 \pi A}} = \frac{1}{2 \pi}\int^{\infty}_{-\infty} d\omega \, \exp\left(i \, \omega \, u - A \omega^2/2\right)
\]
we readily perform the $N-1$-fold integral in the latter equation to get the result in (\ref{Gauss}).

\section*{Acknowledgments}

GO acknowledges helpful discussions with Sid Redner and Julian Talbot. GO
is partially supported by Agence Nationale de la Recherche
(ANR) under grant ``DYOPTRI - Dynamique et Optimisation des
Processus de Transport Intermittents''

\end{document}